# Re-examining the social impact of silver monetization in the Ming Dynasty from the perspective of supply and demand

Tianwei Chang

*Beijing National Day School, Beijing 100039, China*

## *Abstract*

*Existing studies have shown that the monetization of silver in the Ming Dynasty effectively promoted the prosperity of trade in the Ming Dynasty, while the prices of labor, handicraft products and grain were long suppressed by the deformed economic structure. With the expansion of silver application, the fluctuation of silver supply and demand exacerbated the above contradictions. Capital accumulation that should have been obtained through the marketization of labor was easily plundered by the landlord gentry class through silver. This article re-discusses the issue from the perspective of supply and demand. Compared with the increase and then decrease of silver supply, the evolution of silver demand is more complicated: at the tax level, the widespread use of silver leads to a huge difference in the elasticity of production and trade taxes. When government spending surges, the increase in tax burden will be mainly borne by agriculture and handicrafts. At the production level, the high liquidity of silver makes the concentration of social wealth more convenient, while the reduction in silver supply and the expansion of demand have rapidly expanded deflation, further exacerbating the gap between the rich and the poor. Such combined effect of supply and demand factors has caused the monetization of silver to become an accelerator of the economic collapse of the Ming Dynasty.*

\* \* \* \* \*

# 1. Research Background

  Nowadays, China is considered the world's largest producer and is expected to surpass the United States to become the world's largest economy in a decade or so, but this is not the first time that China has demonstrated its economic superpower to the world. Due to its exquisite handicrafts and unprecedented commerce, China had demonstrated great market economic vitality in global trade in the Ming dynasty (1368-1644), and thus became the world black hole for the flow of silver at the time. Depending on the trade surges and the huge demand for currency caused silver produced around the world to flow into China through tribute trade and the Maritime Silk Road. Exports were mainly industrial products such as silk and ceramics, while silver is rarely flown out of China. Thus, silver became China's most important, most in-demand and most valuable single import commodity at the time, and gradually replaced Baochao as the currency of the Ming Dynasty. In 1939, Professor Liang Fangzhong discussed the circulation of silver and foreign trade scale in the Ming Dynasty in his book "International Trade and the Import and Export of Silver in the Ming Dynasty". Taking the Single Whip Law as the starting point of the research, Professor Liang found that the demand for silver in the monetary level in Ming Dynasty China mainly came from government departments[1]. Because the demand for silver is largely determined by the monetization of government taxes. The government actively adjusted the demand in collecting taxes and adopted a large amount of silver. Therefore, after the government obtained a large amount of silver reserves, it used silver for major fiscal expenditures such as national construction projects, military expenditures, and civil servants' salaries. The process of public expenditure eventually laid the foundation for the dominant position of silver in economic activities. Professor Liang further found that although the circulation of silver in the tax system can lead to the development of commodity circulation, it cannot lead to the improvement of handicraft and agricultural productivity. Therefore, in the book "Ming Dynasty Grain Chief System", Professor Liang defined the silver monetization process as "false prosperity"[2]. However, behind the false prosperity, what deeper key factors pushed the economic and social slide to the final collapse in the late Ming Dynasty?

  "Large-scale import of overseas silver" has long been an important topic in the study of the economic crisis in the late Ming Dynasty. Its theoretical logic can prove that there was a relatively serious imported inflation in the middle and late Ming Dynasty at the level of money supply. However, this is not enough to fully explain the formation of the crisis in the late Ming Dynasty. Monetary theory points out that inflation is not an inevitable factor leading to economic collapse. On the contrary, moderate inflation is conducive to the steady growth of the economy. As the scope of research on historical materials continues to expand, scholars such as Zhang Zhongli and Peng Xinwei have proposed that the trend of silver monetization in the Ming Dynasty is not only phased, but its scope of influence does not cover the entire Ming Dynasty: silver monetization in the Ming Dynasty varies in different periods and regions. Therefore, with the expansion of the analytical perspective, the academic community has begun to accept that silver monetization is not the direct cause of the economic recession in the Ming Dynasty, but one of the many factors of the economic recession in the Ming Dynasty. For example, Huang Renyu believes that the economic recession in the Ming Dynasty was caused by the combined effects of factors including rigid social system, rapid population growth, and excessive fiscal burden, and silver monetization is only one of the factors. The extensive analysis of Ming Dynasty historical materials has enriched the perspective of understanding the economic recession in the late Ming Dynasty from a dimensional perspective. However, the academic community has not yet been able to analyze the impact of silver monetization on economic structure and social changes from the overall perspective, especially the changes in the demand side of silver. Since silver in the Ming Dynasty has the dual attributes of currency and commodity, it plays the role of circulation tool, means of production and tax unit in economic activities. Therefore, this paper can apply the supply and demand theory in economics and classify the impact of silver monetization on production, trade and government taxation from the perspective of demand. Re-examine the issue of silver monetization in the Ming Dynasty and its social impact.

---

[1] 梁方仲. 梁方仲经济史论文集[M]. 中华书局, 1989.
[2] 梁方仲. 明代粮长制度[M]. 上海人民出版社, 1957.

  Looking back at the relevant research on the relative price of labor, commodity output and trade in the Ming Dynasty, it can be found that the social productivity in the Ming Dynasty for more than 200 years has not increased significantly, but only accumulated in quantity. At the same time, the monetization of silver has greatly improved the liquidity of the economy and accelerated the accumulation of wealth. This problem of stagnation of production relative to trade was transmitted to the government taxation and fiscal level with the reform of the "One Whip Law", resulting in a situation where the tax elasticity of production-related taxes was extremely small without a significant increase in land. This led to the fact that when wars and natural disasters occurred frequently in the late Ming Dynasty and government spending surged, the increase in tax revenue would be mainly borne by agricultural handicrafts, and the liquidity of the economy would be rapidly impacted. Therefore, the view of this paper is that the monetization of silver is not only one of the key factors causing the social crisis in the late Ming Dynasty, but also an "accelerator" that exacerbated social contradictions in the late Ming Dynasty. Due to the use of silver, most of the surging fiscal expenditure in the late Ming Dynasty was transferred to production-side taxation. At the same time, the economic deflation caused by the insufficient supply of silver was amplified and transmitted to the government through the exhaustion of tax sources, which ultimately led to a simultaneous decline in administrative power and economic vitality. Subsequently, this paper re-examined the impact of changes in the supply and demand of silver on the entire society from the two aspects of commercial activities and social atmosphere. For the first time, based on the demand-side perspective, the multiple roles of silver in the private economy and government taxation were analyzed through this research, and the impact of silver monetization on Ming Dynasty society was further analyzed. The marginal contribution of this paper is to use the economic supply and demand theory and the tax elasticity theory to include the two main aspects of ancient economic activities, production and taxation, into the analysis of the silver monetization issue, which enriched the research perspective of ancient Chinese economic history and provided new ideas for the study of this issue.

# 2. Literature Review

## 1) Concept Definition: What is Silver Monetization?

To study the impact of silver monetization from a supply and demand perspective, we must first clarify the concept of silver monetization. In Chinese history, the use of silver as a medium of exchange began to appear as early as the Tang and Song Dynasties. Gu Yanwu(顾炎武) cited the records of Song Dynasty Renzong, "景祐二年，诏诸路岁输缗钱。福建二广易以银，江东以帛。于是有以银当缗钱者矣。金史食货志记载，旧例银每铤五十两其直百贯。民间或有截凿之者，其价亦随低昂。遂改铸银，名承安宝货。一两至十两分五等，每两折钱二贯，公私同见钱用。" From Gu Yanwu's research, it can be seen that the folk use of silver as a transaction intermediary began to perform the function of money in the Song and Jin Dynasties. In the Yuan Dynasty, the imperial court implemented a treasure banknote system. The treasure banknotes of the Yuan Dynasty were made with silver or other jewelry as reserves, which was the prototype of modern currency. At the same time, physical silver was used in foreign trade. The above measures laid the social foundation for the monetization of silver in the Ming Dynasty. However, in the official documents of the Ming Dynasty, the normative systems related to currency mainly involve "banknote law(钞法)" and "money law(钱法)", while "silver law(银法)" does not appear. This shows that in a legal sense, the Ming court did not impose the same government inscriptions and credit endorsements on silver as it did on banknotes in the early Ming Dynasty. In fact, the silver monetization process was initiated spontaneously by the private sector, while the government was the passive recipient. In the early Ming Dynasty, silver was banned as currency, and the government enforced the "Ming Dynasty Baobao" system. However, due to the lack of precious metal reserves, the Baobao quickly depreciated in value, leading to the failure of the policy. Later, in the transitional stage, during the Hongzhi period of Chenghua, silver was increasingly widely used among the people, so banknotes were gradually abandoned in fiscal revenue and expenditure. Important taxes such as salt tax, tea tax and customs duties began to be levied in silver. Until the Jiajing period, silver became the main currency for taxation and government expenditure, and the tax revenue at various banknotes was gradually changed to silver taels. Finally, in the first year of Longqing (1567), Emperor Muzong of the Ming Dynasty confirmed the legal currency status of silver in legal form, marking the completion of silver monetization. Wan Ming (2007) called the late Ming Dynasty the Silver Age. After systematic research and discussion, he believed that the circulation of silver in the Ming Dynasty began in the Chenghua and Hongzhi periods (1465-1505) and experienced from illegal to legal and bottom-up. The evolution process from top to bottom, and finally popularized; the generalization of silver in the Ming Dynasty promoted the reform of the tax system, promoted the emergence of the market economy and the changes in late Ming society, which became an important symbol of China's social transformation. The generalization of silver in the Ming Dynasty promoted the reform of the tax system, promoted the emergence of the market economy and the changes in late Ming society, which became an important symbol of China's social transformation.

Based on the above research, we can draw a conclusion that the monetization of silver in the Ming Dynasty was not a natural process, but the result of a joint game between the private sector, the market and the government. With the gradual collapse of the treasure banknote system established in the early Ming Dynasty, coupled with the shortage of raw materials for copper coins, private economic activities promoted the trend of silver monetization in terms of currency use. Subsequently, the Ming court took advantage of the situation and gradually unified the use of silver in government expenditures and taxes. Therefore, the so-called silver monetization in the Ming Dynasty refers to the comprehensive process from the evolution of economic flows to government taxation and fiscal expenditures. It is the inevitable result of the combined effects of the nature of currency, private applications, and government guidance.

## 2) Previous Research on "Silver Monetization"

The study of silver monetization was not the first academic research in the 20th century. The impact of the flood of silver on society in the Ming Dynasty had attracted great attention from famous thinkers such as Huang Zongxi(黄宗羲), Gu Yanwu(顾炎武), and Wang Fuzhi(王夫之) as early as the transition from Ming to Qing. Because the flood of silver might cause the reconstruction of social structure and the rapid concentration of wealth, leading to the instability of national rule, all three supported the idea of "abolishing silver". This simple view based on the actual situation has gradually evolved into an important perspective for studying the evolution trend of economic activities in the Ming Dynasty with the enrichment of historical analysis tools since modern times. A large number of scholars started with the analysis of the monetization process of silver and studied the social changes in the middle and late Ming Dynasty, government fiscal and taxation reforms, and production structure. Regarding the causes of the problem of silver monetization, current historical research can be mainly divided into three perspectives: silver inflow perspective, government system reform perspective, and trade perspective. Among them, silver inflow has always been the focus of research because of its detailed number and easy statistics. For example, foreign scholars such as Atwell, Flynn and Giraldez have studied the inflow of overseas silver into China and its impact. Based on the perspective of silver inflow, Quan Hansheng(全汉升) has conducted in-depth research on Taicang silver, international trade and silver imports in the Ming Dynasty, as well as the impact of silver on the northern border and market activities. The inflow of silver into China in the Ming Dynasty had a significant impact on domestic prices and inflation. Peng Xinwei(彭信威) conducted in-depth research on the problem of silver flooding in the Ming Dynasty from the perspective of monetary history, and Ye Shichang(叶世昌) conducted in-depth research on the problem of silver flooding in the Ming Dynasty from the perspective of monetary history, and Ye Shichang conducted in-depth research on the problem of silver flooding in the Ming

Dynasty from the perspective of monetary theory evolution. Peng Xinwei (1994) and Ge Zhaoguang(葛兆光) (1996) both believed that the import of silver led to an increase in the money supply, which led to a general increase in prices, especially the prices of luxury goods and land. However, Peng Xinwei (1994) believed that the impact of inflation was not the same in all sectors, and the prices of basic commodities and wages often lagged behind the overall price level. The perspective of government system reform discussed the issue of silver monetization from the perspective of the Ming Dynasty government's policy changes on currency, fiscal and tax policy reforms, and other aspects. Ge Zhaoguang(葛兆光) (1996) found that the Ming government actively encouraged the use of silver as a tax payment method because it was more reliable and efficient than traditional physical payments. Huang Renyu (黄仁宇)(2002) pointed out that the monetization of silver also enabled the government to better control and supervise tax collection because silver was easier to account for than various commodities. In addition, the availability of silver revenue enabled the Ming government to fund its ambitious infrastructure construction, military expansion, and the construction of the Great Wall. Huang Aming (黄阿明)(2010) found that the institutional operating environment of traditional countries should not be ignored when discussing silver after examining the development process of silver monetization in the Ming Dynasty, because the monetization of silver in the Ming Dynasty started from the field of national taxes and services, and the needs of national finance prompted the emergence of silver monetization. The study of silver monetization from the perspective of trade discusses silver monetization from a broader global perspective. For example, for the rise of the Western world at the same time, Acemoglu et al. (2005) pointed out through empirical research on trade between the East and the West that the direct cause of Europe's rise was the Atlantic trade, and the prosperity of trade activities provided sufficient material guarantee for the bourgeois revolution in the Western world. The prosperity of maritime trade is inseparable from the sufficient supply of precious metal currency. Following this logical thread, Zhang Yuyan(张宇燕) and Gao Cheng(高程) (2004) proposed a unique theoretical model of "currency shock-institutional change-economic growth", pointing out that the greatly increased supply of gold and silver in the Americas led to the emergence of a "price revolution" and the redistribution of social wealth, which led to the rise of emerging merchant groups and the weakening of feudal aristocrats, and ultimately gave birth to capitalist institutional innovation and the rise of the Western world. This explanation proposes that currency shock is the cause of wealth redistribution, which leads to the rise and fall of class power, the disintegration of old social classes, the emergence of efficient economic organizations and the rise of emerging merchant classes. Xiao Fang(萧放) (1989) discussed the relationship between silver and the state's taxes, commercial capital, and social and economic development in the Ming Dynasty from a macro perspective. Wan Ming(万明)(2004) discussed the relevant issues and impacts of silver monetization in a more systematic and complete manner, and believed that silver monetization began in the Chenghua and Hongzhi periods of the Ming Dynasty (1465-1505) and ended in the Jiajing and Wanli periods (1522-1566). The completion of silver monetization was a double-edged sword for the late Ming society. From a positive perspective, the birth of the silver-money system with silver as the main currency and money as the auxiliary currency can avoid inflation caused by the increase in the supply of credit currency (i.e. paper money). Silver, as a precious metal currency, can automatically adjust the amount of currency in circulation according to economic development. Subsequently, Li Xiaotong(李小彤) et al. (2015) explored the factors that led to the overall stagnation of China's development in the Ming Dynasty from the perspective of consumption, and summarized the theoretical model of monetary shock-consumption pattern change-institutional change to explain the economic development of the Ming Dynasty and the role of silver in it. A review of previous studies on the issue of silver monetization shows that as the research on silver monetization continues to enrich reference historical materials, its research foundation is no longer based solely on history, but has introduced theoretical perspectives from multiple disciplines including money supply theory, trade theory, fiscal theory, and consumption theory, analyzing the historical process and scope of influence of silver monetization from multiple perspectives. One of the more typical paradigms is to use economics or finance theory to further analyze the issue of silver monetization. Based on this finding, this article attempts to use the supply and demand theory to reanalyze the issue of silver monetization, in order to discover some new impact mechanisms of silver on the economy and society of the Ming Dynasty and draw innovative conclusions.

# 3) Re-discuss the Theoretical Basis of Silver Monetization

"人主操富贵之权，以役使奔走乎天下，故一代之兴则制之，一主之立则制之，改元则制之，军国不足则制之，此经国足用之一大政也"[3] Jin Xueyan(靳学颜), a famous official in the Ming Dynasty, clearly discussed the importance of the state's control of currency. The reason why silver became the most important form of currency in the middle and late Ming Dynasty is related to its many important properties as currency. First, silver is highly substitutable. Currency is a typical substitute. The same type and amount of currency can be substituted for each other. This characteristic makes currency a universal commodity. In transactions, there is no need to consider the individual characteristics of specific currencies. Since the adoption of silver by the people in the Ming Dynasty, it is difficult to find other commodities that are more substitutable than silver within the Ming Empire. The second is the nature of consumer goods.

From the perspective of civil law, currency is a legal consumer. Although currency is not destroyed after use, for the holder, once it is used, the ownership is lost, which is similar to the nature of consuming other consumer goods. The third is liquidity. The main function of currency is to serve as a medium of exchange and a means of payment, so it must be tradable. This requires that currency must be a movable item with real liquidity. The fourth is the particularity of ownership. The right to possess currency means full ownership, which makes people willing to accept silver currency as a means of payment. From the basic functions of currency, silver perfectly fulfills the four major functions of medium of exchange, unit of account, value storage and deferred payment standard. As a medium of exchange, silver is easy to carry, divide and identify, which greatly facilitates the transaction process. Its relatively stable value makes it an ideal unit of account, providing a reliable value reference for other commodities. Silver's excellent corrosion resistance ensures its long-term value preservation ability and meets the demand for value storage. At the same time, its value stability also makes it a reliable deferred payment standard. From the perspective of the properties of ideal currency, its scarcity stems from its relative scarcity in nature and the difficulty of mining, which guarantees its value. In addition, silver has stable chemical properties, is not easy to oxidize and corrode, has excellent durability, and can be precisely divided into small units to meet the needs of transactions of different scales. The homogeneity of silver ensures that it has the same characteristics everywhere, which is convenient for standardization. The high value brought by high density makes it easy to carry. In addition, the unique color and weight characteristics of silver make it easy to identify authenticity. From the perspective of money supply theory, the natural supply limit and automatic adjustment mechanism of silver ensure its stability as a currency. Silver mining is limited by natural conditions, and the supply growth is slow and relatively stable, which avoids inflation caused by excessive money supply. At the same time, when the supply of silver increases, its value will decrease slightly, which will inhibit the enthusiasm for mining and form a self-regulatory mechanism. The theory of money demand explains people's preference for silver currency. With the development of the commodity economy, silver meets people's demand for a generally accepted medium of exchange. Its value stability makes it an ideal means of wealth storage and meets precautionary needs. And its value fluctuations also provide the possibility of speculation and increase its attractiveness. Although there was no monetary policy in the modern sense in ancient times, the silver standard provided a basis for monetary policy. Under the silver standard, the money supply is equivalent to the silver reserve, forming an automatic stabilization mechanism. The theory of currency evolution, the theory of international monetary system and the theory of currency substitution can analyze and explain the rise of silver currency from the perspective of monetary banking theory. As a rare and useful commodity, silver naturally becomes an ideal medium of exchange. Its physical properties make it the best choice of metal currency. Subsequently, the government intervened in silver casting and further standardized silver currency. The theory of international monetary system explains the importance of silver in international trade: silver provides a unified pricing standard for international trade, and the fixed exchange rate system based on silver provides a stable environment for international trade, while facilitating international debt settlement. The theory of currency substitution explains why silver is widely accepted in different cultures and regions. As more and more people accept silver as currency, its use value is further enhanced, forming a network effect. The physical properties and intrinsic value of silver enable it to transcend cultural differences and be widely recognized. In summary, silver not only perfectly meets the basic functions and attributes of currency, but also shows unique advantages in terms of currency supply, demand, policy, credit, etc. Its scarcity and stability guarantee the value of currency, its physical properties are easy to use and preserve, and its universal acceptance promotes the development of trade. In general, the theoretical basis of silver as currency includes its legal basis, physical properties, economic attributes and historical traditions. These factors interact with each other to create the unique monetary nature of silver in the Ming Dynasty and also construct the underlying logic for the success of silver monetization. The rise of silver currency is the result of economic development, technological progress, government planning and social needs in the Ming Dynasty. Silver monetization reflects the social demand for stable and reliable currency since the mid-Ming Dynasty and the government's plan to simplify taxation and coinage in order to improve administrative efficiency.

---

[3] 明·靳学颜：《皇明经世文编》卷299

# 3. A direct reflection of the monetization of silver

## 1) Monetary Level: The Substitution Effect of Silver

Looking back at the history of the development of Chinese currency, Zhang Zhongli (张仲礼)(2016) believes that before the Ming Dynasty, China's economy was mainly based on copper coins, supplemented by iron, tin, a small amount of gold and silver, and later paper money. Ge Zhaoguang(葛兆光) (1996) found that the influx of silver from the New World of America and other regions in the 16th century led to China's gradual use of silver currency. In stark contrast to the prosperous development of the market economy in the middle and late Ming Dynasty in the 16th century, the monetary economy and market economy in the early Ming Dynasty were extremely shrinking. Under the design of the founding emperor Zhu Yuanzhang, the economic structure of the early Ming Dynasty was far more conservative than that of the Yuan Dynasty and the Song Dynasty. A large number of people were divided into military households, service households, and artisan households, and were forced to participate in large-scale reclamation and migration, and could not participate in the market economy normally. Zhu Yuanzhang hoped that all citizens would live according to the small peasant economy system he built, and the idea of "agriculture-based" greatly restricted the development of the market economy. The backwardness of this economic system is not only reflected in the social division of labor, but also directly reflected in many aspects such as private trade, fiscal taxation, and currency issuance[4]. As the first credit currency in the Ming Dynasty, the Ming Baochao had no fixed issuance limit and no clear issuance reserve. Therefore, by the middle of the Ming Dynasty when the market was prosperous, the government could no longer control the choice of currency. The spontaneous economic law chose silver currency, and the non-convertible credit paper money Baochao was replaced by precious metal silver. The government's unrestrained printing of Baochao caused inflation to soar, and the result of the sharp decline in people's demand for silver could only be the collapse of the Baochao system. Silver had a strong substitution effect on other general equivalents such as Baochao and copper coins. In the 23rd year of Hongwu (1391), the face value of Baochao quickly depreciated to 1/4 of the original regulations; in the 30th year of Hongwu (1398), although the exchange rate between banknotes and silver was still 5 to 1, the exchange rate between Baochao and gold had depreciated to 70 to 1. Although the government ordered the prohibition of gold and silver transactions, the depreciation of Baochao had already shown an irreversible trend. From Yongle to Xuande (1403-1435), there was a transition to physical transactions in silver. In the early years of Yongle (1403), Baochao depreciated repeatedly. In the second year of Yongle, Chen Ying, the imperial envoy, believed that the reason for this situation was that "朝廷出钞太多，收敛无法，以致物重钞轻". In order to maintain the credit of Baochao, the Ming government promulgated the "Household Salt Law", which was completely based on paying with banknotes in principle. In the early years of Xuande, an order was issued: "驰布帛、米麦交易之禁，凡以金银交易者...罚钞", but private transactions were still conducted with gold and silver. Although "易用银一钱者，罚钞千贯；脏吏受银一两者，追钞万贯，更追免罪钞", it still led to "钞滞而不行". "In the transaction of hundreds of things, the only thing held was trust. Once the trust was lost, the people would not follow it." In order to maintain the purchasing power of Baochao, the issuance of new banknotes was stopped in the third year of Xuande (1428), and the rotten banknotes were destroyed; taxes were increased, and taxes were paid with Baochao, and salt was paid with banknotes, etc., to expand the scope of Baochao's use. But by the 14th year of Jiajing (1535), one thousand strings of Baochao could be converted into four silver coins, or 276 coins. The trend of Baochao devaluation was irreversible. Copper coins were also used in parallel in the economic activities of the Ming Dynasty, but they were limited to circulation in some areas. Xie Zhaoshuan sorted out the media used in the exchanges between the north and the south in the Ming Dynasty, pointing out that "the currency and silver ear are the common media in the exchanges in the world today. Using money is convenient for the poor." The currency law in the Ming Dynasty was relatively loose, "the currency law was discussed in the morning and changed in the evening, and there was no final rule." The minting and issuance policies of copper coins were mixed and inconsistent. Whenever the reign was changed, they were minted and used according to the reign name. The Ming government stipulated that "the Jiajing coin was seven coins, the Hongwu coins were ten coins, and the previous coins were thirty coins, which were equivalent to one cent of silver", and in the Jiajing period, "eight coins were equivalent to one cent of silver". In the sixth year of Longqing (1572), "Eight coins on the back of a gold coin are equivalent to one cent of silver, and ten coins on the edge of a wax coin are equivalent to one cent of silver. The Hongwu coins and the old coins of the previous dynasties are equivalent to one cent of silver each." In the 46th year of Wanli (1618), "Nanjing silver is ten coins per cent." In the late Chongzhen period, "one thousand silver coins are worth no more than one coin and two cents, which is not as good as the price of copper." In the second year of Shunzhi (1645), "seven coins are equivalent to one cent of silver, and the old coins are equivalent to one cent of silver at fourteen coins." As the government was unable to control private minting, the quality of coins also declined. The competition between privately minted copper coins and state-issued coins made the government's currency system unworkable, and the quality of privately minted counterfeit copper coins was poor. Under the influence of market rules, people tended to use silver with stable value and good credit as circulating currency. In the first year of the Longqing reign (1567), Emperor Muzong of the Ming Dynasty ordered that "for goods of the people worth more than one coin of silver, silver and coins can be used together. For goods worth less than one coin, only coins can be used." The issuance of this decree marked that silver was officially recognized as a legal currency by the state. Moreover, Yang Duanliu (2007) believes that from the perspective of currency circulation over the past few hundred years in the Ming and Qing dynasties, the importance of silver quickly surpassed that of copper coins. In the case of a huge trade deficit, a large amount of silver flowed into the country in the Ming Dynasty, stabilizing the silver-coin exchange rate, satisfying the monetary needs of the Ming Dynasty's social economy, and bringing great impacts on the development of the social economy. The process of silver popularization is actually a process of competition between silver and treasure notes and copper coins. After the mid-Ming Dynasty, under the conditions of an open economy, there were multiple convertible currencies such as silver, treasure notes and copper coins in circulation in the country. Due to the excessive issuance of treasure notes, there was a serious inflation, and people had weakened their confidence in the stability of the currency. Considering the opportunity cost and relative benefits, people reduced their holdings of treasure notes and copper coins and increased their holdings of relatively high-value non-official currency silver. As a result, silver currency gradually replaced legal tender, namely, treasure notes and copper coins, as a means of storing value or as a medium of exchange. Silver first replaced legal tender from the perspective of value storage, and then gradually replaced legal tender from the perspective of medium of exchange. The "substitution effect" of silver monetization reflects the typical good money driving out bad money, which is in line with the real demands of private economic activities in the Ming Dynasty.

---

[4] 管汉晖, 李稻葵. 明代 GDP 试探[C]. 清华大学中国与世界经济研究中心研究报告(总第 6 期). 2008.

## 2) Government Level: Loss of Seigniorage and Weakened Fiscal Capacity

From the perspective of seigniorage, in the early monetary system where Baochao was the main currency in circulation, the government completely owned the seigniorage. However, under the protection of the "compulsory authority" of the Ming government, the Baochao system eventually could not escape the fate of collapse. Zhao Yifeng (1985) pointed out that the development of the status of silver currency was first reflected in the transformation of the tax system marked by the collection of "gold flowers and silver" in the first year of Zhengtong (1436). The fiscal system shifted from collecting mainly in kind to collecting mainly in currency, which was a reflection of the changes in the natural economic foundation of society. On the basis of this fiscal system, the state needed to strengthen monetary management to achieve fiscal balance. However, the Ming government was unable to grasp and control the circulation of silver, and the central government's ability to control the economy was weakened, which triggered a major fiscal crisis. The fundamental reason why the Ming government tried its best to maintain the Baochao system was that under this system, the Ming government could obtain seigniorage at a very low cost by issuing Baochao, thereby obtaining the greatest benefits in economic circulation. After the monetization of silver, except for the use of lead instead of silver when casting silver coins, there was basically no income from the issuance of currency. In the middle and late Ming Dynasty, the government followed the market situation and established the currency status of silver because of the spontaneous circulation of silver among the people, the wide recognition of the people, and the relatively stable inflow of foreign silver. These conditions objectively helped the Ming Dynasty government to survive a financial crisis. However, when silver was the main currency in circulation, the government hardly received any seigniorage from issuing currency. Therefore, the process from preventing the generalization of silver to reluctantly recognizing the dominant position of silver currency highlights the government's helplessness in the generalization of silver currency, which once again reveals from one side that the legalization and generalization trend of silver currency are unstoppable. However, the silver currency system also contains huge risks, because the supply of silver currency is not dominated by the Ming Dynasty government, and the government did not receive any seigniorage from issuing currency. The "One Whip Law" reform in the ninth year of Wanli stipulated that taxes and labor service were unified, taxes were calculated according to mu, and all taxes were paid in silver. Taxes and labor service were changed to silver. All obligations and responsibilities of the people to the court were converted into silver for delivery. "Silver began to be the only important thing in the world, and all things were based on silver." At this point, under the guidance of the goal of increasing national fiscal revenue, the imperial court's decree was implemented, and the monetization of silver was finally completed. China's land tax system has entered the stage of monetary tax from the stage of in-kind tax. Land tax is the economic source of survival for the courts and government offices of all dynasties, that is, the current government, and it has an extremely important position. Silver naturally became the lifeline of the social and economic life of the Ming Dynasty. The "History of Ming Dynasty" records that "in the winter of the 22nd year of Yongle, in October, Renyin, the gold and silver of the people were abolished, and the two capitals of the Ministry of Revenue were abolished." Silver became "a thing that the state has given the ability to pay" after it "became a medium for general transactions through the actions of people participating in commercial transactions." If the commercial society decides to adopt a new medium, within the scope of the legal rights of both parties to the transaction, it will strive to make it the standard of deferred payments at the same time, so as to deprive the government of the effectiveness of things that are fully repayable; at least deprive it of its future binding effect. From the perspective of government fiscal capacity, before the "One Whip Law" reform, Taicang's annual revenue was 2,014,200 taels of silver, but it increased year by year after the reform, and in 1585 it increased to 3,676,000 taels, an increase of more than one million taels compared to before the reform. In 1587, Taicang's annual revenue was nearly 4 million taels of silver, nearly double the amount before the "One Whip Law" was implemented. In 1592, Taicang's annual revenue increased to 4,723,000 taels of silver within a year, an increase of nearly 3 million taels compared to before the reform. It can be concluded that the "One Whip Law" brought Taicang an annual revenue of 21,313,700 taels of silver in just seven years. When silver is used in official revenue and expenditure, its legitimacy is naturally established. The Ming Baochao was completely abandoned by the people and the market, and the difficulty and cost of the government's extraction of national finances increased greatly, resulting in a significant reduction in the government's exclusive enjoyment and control of finance and related interests. The central government could no longer effectively regulate the social economy and could only rely on political control. Therefore, in the more than 100 years of the middle and late Ming Dynasty, Baochao could only continue to be used as a government bond for some purposes. In the middle and late Ming Dynasty, the Ming government intertwined fiscal policy with monetary policy, and most of the policies adopted were fiscal policies to increase government fiscal revenue. In the process of fiscal game between the central and local governments in the late Ming Dynasty, there was a situation where "few officials paid the two taxes, and more officials paid miscellaneous taxes." The tax and service reform centered on the Single Whip Law created conditions for local governments to expand their fiscal autonomy, but also weakened their motivation to collect and pay the state's regular taxes. In addition, another important principle of the Single Whip Law - the unified collection of taxes and service silver, also led to the aggravation of tax arrears. In addition to the price factors mentioned above, there is also the emergence of "fire consumption", which provides more space and opportunities for local governments to increase their fiscal autonomy and for local officials to embezzle public funds. In the process of collecting and paying fire consumption, local officials arbitrarily increased the consumption rate, which directly increased the burden on taxpayers and aggravated the hidden dangers of tax arrears.

# 4. From loose to tight: Change in silver provision

## 1) Domestic Mining and Silver Reserves

From the perspective of money supply, silver in the Ming Dynasty mainly came from self-production and internal flow. Among them, self-production is divided into domestic mining in the Ming Dynasty and silver reserves from the previous dynasty. Therefore, a large number of scholars have analyzed the overall circulation scale of domestic mining and silver reserves in the Ming Dynasty from the two aspects of production and accumulation in the previous dynasty. Wang Yuxun (1998) examined the amount of silver mined in the Ming Dynasty. Because the Ming Dynasty always strictly adhered to the policy of "(private) silver mines should not be opened lightly", it is more appropriate to say that "the total amount of domestic silver mined during the Ming Dynasty was about 20 million taels." Ge Zhaoguang (1996) believes that the increase in the supply of silver, coupled with the Ming government's policy of encouraging silver imports and accepting silver as tax, has made silver the main medium of exchange. Quan Hansheng (1991) calculated the average annual silver tax of the Ming Dynasty based on the silver tax data in the "Ming Shilu", and estimated that the cumulative silver production from 1390 to 1520 was 309.47 million taels, with an average annual silver production of 300,000 taels. Liu Guanglin (2011) examined the economic development of the Ming Dynasty from the perspective of currency stock. The demonetization policy in the first century of the Ming Dynasty caused the social and government demand for transaction media to fall to the lowest point since the 8th century AD, and the severe shortage of currency also became a major obstacle to economic activities after 1500. Gu Xiuting (2018) collected and sorted out the data on the silver production of the Ming Dynasty in the "Ming Shilu" and found that the amount of silver mined during the Yongle and Xuande years was relatively large, about 200,000 to 300,000 taels per year. After that, the silver production tended to decline, about 100,000 taels per year. Taking the silver production data from the "Ming Shilu" on the Ming Dynasty from the 23rd year of Hongwu to the 15th year of Zhengde and Wang Yuxun's silver production data from the Jiajing Dynasty to the Wanli Dynasty, the silver production of the Ming Dynasty was finally estimated. It can be concluded that the total silver production of the Ming Dynasty from 1390 to 1620 was 22 million taels. During this period, there were 211 years of recorded silver production, and it can be estimated that the average annual silver production in the Ming Dynasty was 105,000 taels. The estimates of the above scholars are roughly the same, estimating that the average annual production in the Ming Dynasty was between 100,000 and 300,000 taels, and the overall mining trend was first increasing and then decreasing.

According to the relevant research data statistics in Liu Guanglin's "Research on the Currency Problem in the Ming Dynasty - Preliminary Estimates of the Scale and Structure of the Monetary Economy in the Ming Dynasty" and Wan Ming's "Monetization of Silver in the Ming Dynasty: A New Perspective on the Connection between China and the World", at that time, China's finances, long-distance trade, and grassroots markets all had a large demand for silver. From 1528 to 1571, the Taicang Silver Treasury in the Ming Dynasty was basically in a state of deficit. However, this situation began to improve after 1567. After 1567, the gap between the annual income and annual expenditure of the Taicang Silver Treasury in the Ming Dynasty narrowed significantly, and even in 1577, the annual income exceeded the annual expenditure, and this period was the period when a large amount of American silver flowed into China. Before American silver flowed into China, Japan had already begun importing large amounts of silver to China in the 1540s. It can be seen that silver flowing from overseas was the main source of silver supply. Therefore, this article will further analyze this.

## 2) Overseas Inflow and Total Estimation

The monetization of silver in the Ming Dynasty was a major turning point in the history of Chinese currency. According to the theoretical analysis of supply and demand, in terms of supply, the substantial increase in the global silver supply since the 16th century was the material basis for the realization of silver monetization in the Ming Dynasty. As mentioned above, the increase in silver production and imports from Japan and other places, coupled with the inflow of silver from the Americas, greatly increased the available amount of silver. At the same time, the development of maritime trade and the establishment of the "tribute trade" system broadened the circulation channels of silver and promoted the widespread circulation of silver in East Asia.

The exogenous supply of silver is divided into two categories: inflow from Japan and inflow from the Americas, and analyzed separately: Qian Jiang mentioned in his article "An Investigation of International Silver Flows and Its Import into China in the 16th-18th Centuries" that "According to the official investigation of the Japanese shogunate, from 1601 to 1764, a total of 4,212,295 kilograms of silver were exported from Japan, most of which flowed into China. According to Atwell According to Li Longsheng's view, based on the estimate that 1 ton is equal to 26,600 taels of silver, it can be concluded that Japan's annual silver production from 1560 to 1600 was 1,330,000 taels, and from 1601 to 1640 it was 3,990,000 to 5,054,000 taels. Then, it can be inferred that Japan's total silver production from 1560 to 1640 was 212,800,000 to 255,360,000 taels. Li Longsheng believes that Japan's total silver production from 1560 to 1644 was 254.29 million taels, which is not much different, and the average annual silver production was 2,240,000 taels. Regarding the issue of Japan's silver production, the views of experts and scholars are not much different. Wu Chengming believes that the amount of silver flowing into China from Japan from 1540 to 1570 was 7.5 to 15 million taels, from 1570 to 1600 was 15 to 18 million taels, from

1600 to 1630 was 30 to 45 million taels, and from 1630 to 1647 was 14.4 million taels. Based on these data, we can conclude that the total amount of silver flowing into China from Japan through direct trade between China and Japan from 1540 to 1647 was approximately 66.9 to 92.4 million taels. From 1648 to 1672 was 18.48 million taels, from 1673 to 1684 was 5.39 million taels. From 1540 to 1700, the total amount of silver flowing into China from Japan through direct trade between China and Japan was 112.87 million to 138.37 million taels. The amount of silver flowing into China through direct trade between China and Japan was about 100 million taels, and the amount flowing into China through re-export trade was about 80 million taels. Therefore, the total amount of silver flowing into China from Japan was about 180 million taels. The views of scholars such as Peng Xinwei, Zhuang Guotu, and Li Longsheng are not much different. Therefore, this article mainly uses this data to estimate the silver reserves in the Ming Dynasty.

  Regarding the amount of silver flowing into China through Sino-Philippine trade, different opinions vary, and There is a big gap. Wan Ming believes that the total amount of silver flowing into China from 1570 to 1644 was 203.2 million taels, which is much higher than other scholars. His estimate may be too high. Liang Fangzhong believes that the amount of silver flowing into China from 1573 to 1644 was 20.45 million taels, which is lower than other scholars. His estimate may be conservative. In order to estimate more accurately, we combine the views of Wu Chengming in Table 35 and take the average of each. It can be concluded that the amount of silver flowing into China through Manila in the Ming Dynasty was about 70 million taels. Wu Chengming believes that the amount of silver flowing into China through Sino-Philippine trade each year from 1570 to 1579 was 285,000 taels. It can be concluded that the total amount of silver flowing into China from 1570 to 1579 was 2.565 million taels. By analogy, the total amount of silver flowing into China from 1580 to 1589 was 8.001 million taels, 6.327 million taels from 1590 to 1599, 9.369 million taels from 1600 to 1609, 9.333 million taels from 1610 to 1619, 8.109 million taels from 1620 to 1629, 12.582 million taels from 1630 to 1639, and 6.192 million taels from 1640 to 1649. Therefore, the total amount of silver imported into China from 1570 to 1649 was 62.478 million taels.

  In addition to the above-mentioned route of American silver flowing into China through Sino-Philippine trade, another route was through Spain, Portugal, Britain, the Netherlands and other countries. In order to give us a more intuitive understanding of the amount of American silver flowing into China through different European countries, we collected relevant research data from Hou Zhigang's article "A Study on the Inflow of Foreign Silver into China (16th-mid-19th Century)". According to the relevant research data in Wang Yuxun's article "A Trial Study on the Domestic Mining and Foreign Inflow of Silver in the Ming Dynasty" and Li Longsheng's article "Estimates of Silver Stocks in the Late Ming Dynasty", in the late Ming Dynasty, a total of 87,750,000[2] taels of Spanish silver flowed into the Ming Dynasty, a total of 200,000,000[3] taels of Japanese silver flowed into the Ming Dynasty, and a total of 42,762,750[4] taels of Portuguese silver flowed into the Ming Dynasty, totaling 330,512,750 taels. According to Von Glahn's point of view, the amount of silver flowing into China from Europe in the Ming Dynasty was about "32.8 million taels", and the amount of silver flowing into China through Sino-Philippine trade was about 70 million taels. Combining the views of various parties, it can be roughly estimated that the amount of silver flowing into China in the Ming Dynasty was about 330 million taels. Regarding the amount of silver in China before the Ming Dynasty, Li Longsheng believes that "the silver production of the Tang, Song and Yuan dynasties was about 377.75 million taels". In addition to the foreign silver inflow, we estimated earlier, which was about 300 million taels, and the total domestic silver production was about 29 million taels, we can roughly estimate that the total supply of silver in the country during the Ming Dynasty was about 706.75 million taels.

    From the trend point of view, except for a decline in 1590-1599 due to the Japanese invasion and the sea ban, the supply of silver was generally in an upward state between 1570 and 1639. In the late Ming Dynasty, from 1639 to 1649, the silver supply fell rapidly. The hidden dangers brought about by this impact will be analyzed in detail below. In general, the change in the silver supply in the Ming Dynasty went through a complex process. This change was deeply rooted in the specific economic, social and political environment at that time. The fact that the silver supply first increased and then decreased not only reflects the important impact that global trade had on the Central Plains dynasty for the first time, but also reflects the driving force of institutional changes and government reforms within the Ming Dynasty. It is also closely related to changes in the international economic landscape.

# 5. Top-heavy: Changes in silver demand

## 1) Trade Level: the core driver of the silver tide

Since the mid-Ming Dynasty, on the one hand, the booming commodity economy has created a huge demand for universal equivalents. On the other hand, the acceleration of urbanization and the rise of the merchant landlord class have given rise to the whole society's demand for assets that are easy to store and trade. The superposition of the two demands has made the monetization of silver more deeply penetrated, contributing to the developed market economy network in the Ming Dynasty. The development of market transactions, in turn, further stimulated the demand for silver. Eventually, the government gradually accepted silver as a tax and transaction medium. As the acceptance of silver currency by all social classes continued to increase, the demand for silver was greatly strengthened in the Ming Dynasty, so much so that Gunder Frank believed in "Silver Capital: Focusing on the East in Economic Globalization" that "the wheels of the global market from 1400 to 1800 were lubricated by the global flow of silver." The essential reason for the evolution of silver from a precious commodity to a currency is also the development of commerce in the Ming Dynasty. "As long as there is a commodity economy, the more detailed the division of labor and the higher the human desire, the more urgent the need for indirect exchange." In addition, Wan Ming (2004) concluded that the key turning point of silver monetization in the Ming Dynasty occurred during the Chenghua and Hongguang periods (1465-1505). The frequency of silver-commodity exchange increased further, and it had multiple meanings such as realizing commodity prices and replacing commodity values. At the same time, the types of commodities exchanged with silver increased further, and the silver obtained from the sale of commodities was used to purchase commodities, forming a circulation structure of "commodity-silver-commodity". In addition to commodity transactions, the "silver-valued silver" model also appeared in the loan relationship. Besides, the use of silver reduced restrictions on the circulation of cross-regional financial flows, reduced circulation costs, and even the phenomenon of silver circulation through foreign exchange.

Except the formal trade, during the Ming Dynasty, due to the overseas trade policy that was sometimes open and sometimes banned, smuggling activities also became an important part of Ming Dynasty trade. In the early days, the Portuguese who established a trading post in Macau were the main targets of smuggling trade. According to Petro Maffei, "The Chinese sell everything but buy nothing." Subsequently, the Chinese's desire for silver also left a deep impression on trade explorers from the Spanish, Dutch and other European countries. According to the Spanish Sebastian Manrique, "The Chinese went to hell to find new goods in exchange for the rials they craved (a silver coin popular in the Spanish colonial era, known for its high silver content). They even said in stuttering Spanish, "plata sa sangre" (silver is blood)." He also concluded that at any time, any country that wanted to trade with the Chinese had to pay entirely in silver. People at the time concluded that "without silver, no business from Nagasaki to Surat could be successful." It can be seen that the desire of Ming Dynasty merchants for silver had become one of the main driving forces of global trade at that time.

    As the degree of participation in exchange continued to deepen, silver became a tool for realizing the transfer of commodity value and a medium of universal exchange in the Ming Dynasty. All regions, fields, and classes of Ming Dynasty society were more widely involved in the silver expression system, and exchanges increasingly broke through regional restrictions. The generalization of silver currency has achieved unprecedented development. Huang Renyu (2002) believes that the monetization of silver provides a stable and universally accepted unit of account, which promotes domestic and foreign trade. Peng Xinwei (1994) further pointed out that the monetization of silver enabled the integration of regional markets and silver became a national currency. This integration in turn promoted the development of commercial networks and the specialization of production.

## 2) Production Level: Increasing Silver Demand leads to Low grain prices, Hurting farmers

    In addition to being a circulation tool, the relatively stable value of silver can also reflect the relative price levels of production factors in the Ming Dynasty. Therefore, the demand changes for silver can be studied from the production level. In the mid-Ming Dynasty, a large amount of imported silver mainly flowed into the production and circulation links. The long-term, mild inflation level driven by external demand was undoubtedly very beneficial to the economic expansion of the Ming Dynasty (Chen Chunsheng et al., 2010). The prosperity of commerce led to the prosperity of the Ming Dynasty's economy, and the gradual rise of the merchant class, but it also exacerbated the widening gap between the rich and the poor. The boom in trade contrasts sharply with the slump in commodity and labor prices. The persistent price downturn has led to insufficient development of the real economy. Low wages make it difficult for people to accumulate capital, and they are on the verge of poverty for a long time. Two scholars, Peng Xinwei and Quan Hansheng, noticed that compared with the Song Dynasty, the prices of goods, labor services, and food in the Ming Dynasty were very low, and the relative price of silver continued to appreciate. From the Ming Dynasty until the first half of the 16th century, the price of rice remained at the level of the early Song Dynasty (an average of about 0.46 taels per shi), which was a huge gap compared with the peak price period of the Song Dynasty (the average price during the Song peak period was 1.8 taels per shi). The price of rice during the peak period of the Song Dynasty was about three times higher in silver terms than in the Ming Dynasty. If calculated

in copper coins, the price in the early Ming Dynasty was roughly 370 Wen per stone, and the peak in the Song Dynasty was 9,000 Wen per stone. It can be seen that the average price of rice in the Ming Dynasty was several times higher than that in the Song Dynasty, whether measured in silver or copper coins. The low prices in the early Ming Dynasty directly led to low labor wages. For example, the salary of bureaucrats and soldiers in Jiading area differed by as much as 15 times between the Ming Dynasty and the Song Dynasty. For example, the salary of ordinary soldiers was about 3,000 Wen a month in the Song Dynasty and about 200 Wen a month in the Ming Dynasty. However, this low labor price level is not based on strong national power, but a direct reflection of the real economy in the early Ming Dynasty.

Under the impact of the silver wave, unlike the hyperinflation triggered in the Western world, Ming Dynasty society experienced significant productive expansion (Wang Guobin, 1999). However, this production expansion is only a quantitative increase and does not significantly improve social productivity. In the late Ming Dynasty, the official handicraft industry gradually declined, and the private handicraft industry began to emerge. With the implementation of the "One Whip Law", the surrendered silver replaced "forced labor" and "corvee labor". The human resources liberated from it promoted frequent economic population flows and further catalyzed the circulation of silver. Comparing the records about agricultural tools in Xu Guangqi's "Tiangong Kaiwu" with the dozens of agricultural tools listed in Wang Zhen's "Agricultural Book", there is almost no difference. It can be seen from this that there was no obvious improvement in agricultural production technology in the Ming Dynasty. The improvement of agricultural production is mainly reflected in people's management and management capabilities of production. There is an essential difference between commercial trade and industry and agriculture. Trade is not productive labor, but an intermediary connecting production and consumption. Commercial trade can generate value by changing the spatial configuration of resources and commodities, but this activity does not essentially generate new material wealth. Therefore, if the prosperity of commercial trade cannot be transmitted to the production link, that is, commercial capital completes the transformation into industrial capital, this will prosperity is unsustainable. At the same time, Venice and the Netherlands, which had risen due to their commercial capabilities, also turned from prosperity to decline.

From the perspective of farmers' annual profits from farming in the Ming Dynasty, He Liangjun's "Si You Zhai Cong Shuo" in the Ming Dynasty recorded that "a husband and wife can plant twenty-five acres, and those who are less diligent can cultivate up to thirty acres. And they can get more soil and fertilizer, and one can harvest three stones per mu." No matter, if you only collect two shi and five dou, you can get 70 to 80 shi of rice per year. "It can be seen that a farmer with high-quality fertile land can get between thirty-five and forty taels of silver every year. Excluding the taxes and rent paid, one acre of land can earn about one stone of rice, which can be converted into about twelve to twenty taels of silver. Comparing with the previous analysis of labor prices, that is, labor costs, we can see that the income of farmers is also relatively low. If the quality of the fields is average or water conservancy is blocked, the income will be further reduced. According to records in the "Book of Supplementing Farmers", landowners can obtain more than half of the harvest from tenants, making a profit of approximately seven taels of silver. In the early years of Jiajing, Zhang Cong wrote in a memorial that the coastal kitchen households were suffering from delinquency and had no silver and were killed: "... the owner of the stove, Yan'er, is now collecting all the zese, and the loan is double the interest, and the ten rooms are Jiukong, often forced to flee, unable to make a living. "Gu Yanwu lived in Lu for a long time in his later years, and witnessed the suffering of the local people without money. "I saw people from Denglai and Haihai saying that the valley was low, and they lived in remote mountains and had no money to lose their official positions. ". Later, he traveled to Guanzhong. The people "had grain but no silver. What they gained was not what they lost, and what they sought was not what they came from." Xiangsui sells his wife." If things go on like this, there will inevitably be a vicious cycle of poorer crops and poorer people, poorer people and less riches. Year after year, arrears will naturally accumulate. It is common for farmers to be forced to sell agricultural products at low prices not only when taxes are imposed, but also in good years because of low prices for grain or rice. Xu Zan, a famous minister during the Jiajing period, once said: "The example of subtracting the capital from the capital may be suitable for places in the north where boating is inaccessible, but not in the south. It may be feasible where the river rice is expensive, but it is inconvenient to go up the river." In Huguang, Jiangxi, and Jiangbei, the price of rice is not very high, and there are only one or two that are worth seven or eight qian in silver per stone, which is uneven, so the people are forced to sell all their rice, which is not what they want. "Xu Jie, a scholar in the cabinet, also said in a letter: "The purpose of paying for grain is to benefit the people. Unexpectedly, when the rice is cheap, it will be even more disturbing to the farmers. One year, the days were in full swing, and they were looking forward to a good harvest, but the price of rice in the market was low. If the money was exchanged after the tax was collected, it would become a problem for farmers. Not to mention the years of famine? It can be said that in most cases, it is very difficult for farmers to make profits from the production of agricultural products.

  From the perspective of total economic volume, even if the use of silver and other currencies is included, the total economic scale of the Ming Dynasty before 1580 was only about 100 million taels of silver, which was only one-quarter to one-fifth of that of the Northern Song Dynasty. This period was almost the most stable period of Ming Dynasty rule. After 1580, continuous civil uprisings, droughts and wars quickly depleted the accumulation of wealth among the people. Therefore, overall, the low prices in the Ming Dynasty actually reflected the severe reality of deflation and the slow development of the real economy. The use of silver has exacerbated these two realistic characteristics. In the late Ming Dynasty, with the import of large amounts of silver and frequent disasters, the price of rice rose rapidly and reached the level of the Song Dynasty. According to the book "Jin Lei Zi", in the 32nd year of Jiajing (1553), there was a great famine in the capital, and the price of rice suddenly rose to 2.2 taels of silver per stone. In the second year of Jiajing's reign, Nanjing was hit by drought and epidemic. Deaths were everywhere, and the price of rice in warehouses was so high that one stone was worth 1.3 taels of silver. In the summer of the 16th year of Wanli (1588), there was another famine, and the price of rice rose rapidly. The price of Japonica rice rose to 2 taels of silver per stone, and the price of warehouse rice was also worth 1.5 taels of silver per stone. However, such high rice prices only lasted for a month or two and then fell quickly. It can be seen that "low grain prices hurt farmers" was not an empty statement in the Ming Dynasty. When discussing the collection of land taxes, Mr. Lu Simian pointed out: "Farmers own grain, and the money they lack is currency. Taxes must collect currency, forcing farmers to exchange grain for currency. The price of grain often falls during the period, and the profits are attributed to the merger." Therefore, Ren Yuanxiang, a scholar of the Qing Dynasty, clearly pointed out the five harms caused by the use of silver in the Ming Dynasty. "Silver cannot be obtained without trade, and it is the third harm if people chase it away." Guo Zizhang also recorded in "Qian Gu Yi" that the grain is cheap and the silver is scarce and the people's living conditions are poor. "Today, compared with the previous years, the sick years are often poorer, and the grains are poorer... Today, compared with the previous years, the sick grains are often poorer, and the grains are poorer. The people are getting hungrier. ...The problem is not that there is too much grain, but that there is too little silver. If there is too little silver, the officials will have to sell it at a low price. If they sell it at a low price, the grain will be depleted and the people will be short of food. As the day declines, the people's lives improve." It can be seen that the acquisition of silver is uneven, especially in some poor areas, which cannot easily obtain enough silver to pay taxes, resulting in a heavier tax burden on farmers than before the use of silver.

## 3)  Taxation Level: One-Whip-Law and Tax Flexibility

In the early Ming Dynasty, the national economic system established by Zhu Yuanzhang was a small peasant economy system based on physical finance. Under this system, economic activities were extremely conservative, and the financial sources of the court and central government agencies mostly came from the direct payment of physical land taxes. The manpower and financial resources required for local government operations and local public affairs and the salaries of civil servants also mainly came from the supply of Lijia. The mechanism for the state to regulate treasure notes was also through physical financial means such as the salt household registration system. This semi-militarized system of material requisition could still operate effectively when the bureaucracy and bureaucracy were relatively streamlined. However, after Yongle and Xuande, the financial operation of the Ming Dynasty became increasingly difficult. In order to adapt to the increasingly complex management needs and the use of silver, at the central financial level, the Ming government established the Taicang Silver Vault in the seventh year of Zhengtong. This move marked the completion of the "silverization" of the Ming Dynasty national system. At the level of the tax system, the Jiajing period carried out a landmark tax reform, namely the "One Whip Law", which merged various types of taxes (grain tax, corvée, etc.) into one item and limited the collection to silver. This move completely established a monetary economic circulation system with silver as the core of circulation.

The "One Whip Law" not only merged various miscellaneous taxes into a single tax paid in silver, but also gave rise to the reform of the local fiscal budget system. When the One Whip Law was implemented, the reform of the tax and service system changed the relationship between the court and the local government, and between the government and the people, giving silver an unprecedented importance in the country's administrative operation. The people paid taxes in silver, allowing the government to realize the monetization of taxes and reduce the collection and management costs of in-kind taxes. In addition, by replacing corvée with silver, farmers no longer had to fulfill the heavy corvée obligations, but paid taxes in silver, which greatly simplified the types of taxes. The standardization and unification of taxes made it easier for the government to conduct fine management and verification and supervision of taxes. The One Whip Law reform, with the standardization of tax quotas and the standardization of tax objects as the center, gradually formed the fiscal budget of prefectures and counties. First, a budget system was formed to assign four types of service (usually collectively referred to as "four services"), namely, equalization, equal labor, civilians, and post stations, and to assign quotas of materials for the supply of tribute. After the implementation of the Single Whip Law, the part of the "four services" collected according to land grain was combined, and then combined with the land tax converted into silver to form the so-called "land silver", and the parts of various services originally collected according to the number of people were also combined, namely the so-called "people silver". Therefore, "people silver" is the result of the implementation of the Single Whip Law. Under this system, silver became the main means of taxation and service: the financial sources of governments at all levels and their financial relations were all based on silver as the basic unit of measurement and payment, so that the originally different taxation items had a unified calculation and payment method, and the county financial system was established, which completely changed the relationship between local governments and the central court in traditional China.

In addition to reconstructing the local tax system, the Single Whip Law also had a huge impact on different types of taxes. According to Huang Renyu's estimate, before 1600, the total value of land tax in the country was slightly more than 21 million taels, and the service silver was about 10 million taels. Wu Chengming estimated that the service silver for the two items of Lijia and Junyao was about 22.8 million taels based on the figures compiled by Tang Wenji. In addition, in the middle of the Ming Dynasty, there was another heavy levy, collectively known as "supply materials", which many counties levied separately from the tax and service. Taking this levy into account, it is completely reasonable to estimate that the tax and service silver in the late Ming Dynasty was as high as 40 million taels. In the fiscal revenue of the late Ming Dynasty, in addition to the two taxes and corvée, there were several important silver currency revenues, such as the salt tax, which was about 1.3 to 2 million taels. The three additional taxes in the late Ming Dynasty reached more than 10 million taels. Even if the Liao tax, which was set as the annual amount in the 48th year of Wanli (1620), was calculated, it was as high as 5.2 million taels. In addition, there were banknote customs, tax supervision, tax collection, corruption and fines, etc. In the seventh year of Xuande, the commercial tax and fish tax in Huguang, Guangxi, Zhejiang and other places were converted into banknotes, and each tael of silver was paid in banknotes of 100 strings. In the first year of Zhengtong, the salt tax of the Ming Dynasty gradually became the most important part of the court's finances after Chenghong. It was also during this period that the salt tax began to change from physical objects to monetary. During the Chenghua period, the salt tax in Liangzhe began to be converted into silver. The household salt and banknote system established in the early Ming Dynasty had also changed to silver currency by the Chenghua period. In the sixth year of Hongzhi (1493), the inner palace transport treasury was short of gold and silver. The court officials gathered to discuss and proposed that "from now on, each banknote will be converted into three cents of silver, and seven coins will be converted into one cent of silver. Those that should be sent to Beijing will be directly deposited into the inner treasury; those that should be kept will be kept in the place where they are used as salaries for officials and soldiers." In the eighth year of Hongzhi, the Sichuan Provincial Administration was ordered to reduce the tea tax from the second year of Hongzhi, and the amount of arrears in subsequent years will be reduced. Customs duties were the most important commercial taxes in the Ming Dynasty. The conversion of customs duties to silver began in Chenghua. The banknote customs was first established in the fourth year of Xuande (1429). As the name suggests, the establishment of banknote customs was for the purpose of increasing tax revenue in order to popularize the banknote law at that time. The Ming Dynasty successively established banknote customs along the canal and Chang Tax offices were set up at the mouths of the Yangtze River and at key points on the southern waterways. As the circulation of silver expanded, fiscal demand grew. In the first year of the Hongzhi reign (1488), due to "empty treasury", the Ming Dynasty issued an order that "except for the Chongwenmen, Shangxinhe, and Zhangjiawan offices, which will continue to collect money and notes, the eight tax offices in Hexi and other places and the seven tax offices in Linqing, Huai'an, Yangzhou, Suzhou, Hangzhou, Liujiajie, and Zhengyang will all collect silver in accordance with their respective regulations." The total amount of the above items is also more than several million taels. Based on this, it can be conservatively estimated that the amount of silver currency absorbed into fiscal operations each year in the late Ming Dynasty was as high as 40 to 50 million taels. In other words, after the mid-16th century, China's state apparatus and bureaucracy depended on the massive inflow of silver from America to function properly. This silver was absorbed into the fiscal operation system through market circulation, and then entered the market circulation through fiscal expenditure and consumption by the state and bureaucracy, becoming a major force driving the market. Frank even believed that from 1400 to 1800, "the wheels of the global market were lubricated by the global flow of silver."

From the perspective of tax elasticity, land tax was undoubtedly the basis of taxation in the Ming Dynasty, and it was the most stable and least elastic type of tax. A major feature of the basic structure of the Ming Dynasty's national fiscal revenue was taxation. The imbalance of content is specifically manifested in the excessive reliance of annual revenue on land tax. During the Ming

Dynasty, land tax accounted for more than 70% to 80% of the total annual revenue. Despite the great changes in the fiscal system level of the "One Whip Law", the characteristic of excessive reliance on land tax for revenue content ran through the entire Ming Dynasty. A report from the Ministry of Revenue in 1502 listed the country's main income items. The most important of these is the land tax, which accounts for about 75% of all tax revenue, and the total annual revenue reaches 26,799,341 shi of tax grain. "In 1377, Zhu Yuanzhang dispatched officials from various ministries, students of the Imperial Academy, and eunuchs to inspect the individual tax bureaus to fix their tax quotas. In 1393, the land tax revenue reached 322,789,900 shi. Zhu Yuanzhang then announced that "newly reclaimed land in the northern provinces will never be taxed. Since then, fixed tax rates in various places have become unwritten laws. Although there have been occasional adjustments later, the basic quotas have not been abandoned. By the Xuande period, the annual land tax revenue had remained at around 27 million shi. The increase in population and the increase in the amount of arable land were not taken into account. New acres of land were rarely reported, which also made the tax revenue inconsistent with the area of arable land. In addition to the land tax, the levy of labor and basic items also adopted a quota system. This tax quota made the tax elasticity extremely low, and the government could only collect almost unchanged tax amounts for a long time, causing many government functions to be unable to maintain. In addition, because the existing tax revenue could not maintain normal expenses, the lack of funds made the management expenses of many departments insufficient, and the salaries of officials did not rise in sync with the price of goods, resulting in widespread corruption among officials in the Ming Dynasty. From the seventh year of Jiajing to the Longqing In five years, the annual deficit of the Taicang Silver Treasury in the 42nd year was more than 1.8 million taels. This huge imbalance in fiscal revenue and expenditure made the Ming Dynasty's rule precarious. In order to make up for fiscal expenditure, new items such as consumption tax were widely used, and at the peak, it even reached more than half of the tax grain. This new tax with high elasticity gradually became a fixed income with the application of silver. However, due to the huge cost of consumption tax, the burden on the people increased. In order to quell public grievances and reduce the loss of transportation, the Ming Dynasty established the gold flower silver system. In addition, the system of converting tax grain into silver and the system of converting labor service into silver has been gradually implemented since the middle of the Ming Dynasty. This is a major progress in the ancient Chinese tax system. A 15% quota was permanently drawn from the land tax income, which made the government's previous gains from the people through "miscellaneous items" standardized and institutionalized, and then became the government's actual income, so that the low elasticity tax revenue obtained by the government increased.

# 6. Analyzing the impact of silver monetization on economy and society

  The connotation of silver demand in the Ming Dynasty includes private production, government taxation and market trade. The demand in these three aspects almost covers all economic activities in the Ming Dynasty. For this reason, the monetization of silver in the Ming Dynasty had a profound impact on the economic activities and social life of the entire Ming Dynasty, and some of the impacts even spread across hundreds of years to the end of the Qing Dynasty. The study of the social impact of silver monetization starts with the level of taxes and services. As a fiscal policy to implement Chinese history, the tax and services -related systems can clearly reflect the process of silver gradually becoming monetized and gaining a dominant position. The monetization of silver in the Ming Dynasty is also the process of a series of tax and services reforms being promoted. As Wan Ming analyzed, the tax and services reforms in the Ming Dynasty showed three major trends: first, the transformation of in-kind taxes into monetary taxes, second, taxes and services can be uniformly replaced by silver, and third, the transformation of head taxes into property taxes. These three trends are closely related to silver. Taking the "One Whip Law" as an example, the tax and services items were simplified and agreed to be levied in the form of currency, and the grain tax began to be converted into corresponding amounts of silver. This measure greatly reduced the cost of taxes and services and improved fiscal efficiency. The reason for the implementation of the "One Whip Law" was mainly due to the drawbacks of the head tax and the frequent land annexation by the privileged class, which led to farmers' tax evasion and the deficit caused by insufficient fiscal revenue. Corresponding measures include land surveying and promoting the transformation of head tax to property tax to alleviate the oppression of civilians and the problem of the gap between the rich and the poor. Of course, after Zhang Juzheng's death, most of his reforms were abolished. The land did not completely replace the population as the tax standard, but became vague and varied from region to region. In addition, the complicated steps of converting silver into taxes and labor service were accompanied by potential losses such as fire consumption and corruption by officials. Whether it reduced the burden on the people is a matter of opinion. But it is certain that with the reform of taxes and labor service, the monetization of silver entered its peak stage.

  With the monetization of silver, people gradually began to free themselves from the constraints of identity and region, breaking away from the restrictions of the natural economy at both the abstract and concrete levels, directly pushing farmers from paying grain to serving as laborers to paying silver instead of serving as laborers, and the relationship between farmers and the state from identity to contract to a new economic structure of commodity market economy. In terms of identity, the unified monetization of taxes and corvée changed the Lijia system that had existed since the Yellow Book System, and tax grain was no longer a rope that tied people to their sole identity as self-cultivating farmers. As the standard for all taxes and labor services and the medium for value exchange, silver as a circulating currency opened up a broad market that was not limited to a single form, and people had more opportunities to choose jobs. Some farmers who lost their land in the mergers were integrated into the newly established contractual relationship as freelance workers. As recorded in Pengxuan Bieji, "There are many poor people in the western suburbs of the capital. They work as hired laborers every morning, earning wages to support themselves, and return home after work." There are also hired laborers in economically developed areas. According to the Gusu Zhi of Zhengde, in Suzhou, "the proletarians go to various places for hired laborers, receive wages and work, suppress their minds and exhaust their strength, and are called busy laborers. If they have little time, they go fishing and shrimping, collecting firewood and bricks, doing hired labor, and carrying loads, and they are unwilling to be lazy." There are still hired laborers in remote mountainous areas. Qiu Jun, a Ming Dynasty scholar, said that after the refugees from Jiangxi came to the Jingxiang mountainous area, some of them "specialized in trade and hired labor", and should be called "livelihood households". Another group of people chose to come to the other end of the contract as capitalists, just as the textile industry that emerged in Jiangnan during the budding wave of capitalism in the mid-Ming Dynasty, and the iron smelting and iron casting industries in Guangdong, it can be seen that capitalists have the power to choose savings and investment. In addition, commerce and handicrafts flourished during this period. The scale of merchants expanded, and silver was used as currency, which complemented the rise of hired workers. "Along the coast, the fields were all barren and salty. Farmers had no hope for a good year. They could only look at the abyss as a hill. It became a habit. The rich could collect goods and bring them back home. The poor could earn a lot of rice by working as hired workers." It can be seen that in the process of silver monetization, the elements of the commodity market economy, such as demand, supply, savings, and investment, gradually emerged, and the natural economy gradually transformed, all stemming from the dominant demand for money. As Marx proposed, from dependence on people to dependence on things. Under the combined force of tax reform and real market demand, the status of silver as currency became higher and higher, until it set off an unprecedented currency craze in Chinese history. In Ming Dynasty society, the market system was increasingly improved, and silver became the only currency in circulation. All value exchanges in social life used silver as a unified measurement standard. This precious metal with the function of preserving value penetrated all levels of society. Officials could use it to pay taxes, and ordinary people could use it for daily consumption or long-term investment. The widespread practicality of silver has inspired a strong pursuit of it by all social classes. The status of silver in Chinese history is unprecedented, marking a peak in the degree of social dependence on currency. This extreme dependence on silver not only reflects its core role in economic activities, but also reveals the great importance that society attached to monetary stability at that time.

  The monetization of silver gave many people the opportunity to break away from geographical restrictions and move out of their villages to do business in a larger area, which to a large extent promoted subtle changes in the social structure. As people began to leave their villages to seek business opportunities across the country, businessmen engaged in related industries established "business gangs" by creating local ties, and mobile business groups were born. In the middle and late Ming Dynasty, business gangs were particularly common and widely distributed, which also led to industry competition. The powerful business gangs began to seek monopoly status by establishing relationships with bureaucrats and the court. For example, the most famous Anhui merchants in the south obtained the monopoly rights of salt through the imperial court and then achieved a monopoly; Shanxi merchants monopolized the entire country's financial industry through the imperial court, and their power continued into the Qing Dynasty. During this period, class mobility also began to occur, and businessmen gained a way to climb into the upper class. However, in the traditional value concept of Chinese society, scholars, farmers, industry and commerce are separated from each other, and the status of scholars and merchants is significantly different. Taking the imperial examination to become a scholar is supposed to be the only way to achieve class mobility. Therefore, with the trend of increasing business status, merchants want to better integrate the upper class began to become arty and gain cultural recognition. This is why wealthy businessmen in the late Ming Dynasty hired craftsmen and hired workers to build garden sculptures at high prices. As described in "Taicang Gu's Residence" in the 14th year of Chongzhen in the Ming Dynasty (1641): "The rich and powerful family spend their days in the garden pavilion, flowers and stones." For entertainment, but exhausting resources and resources to provide a lot of support..." This is called the merging of scholars and merchants. Changes in moral concepts are contained in changes in social atmosphere. As mentioned above, economy and commerce have gained unprecedented importance with the increasing credibility of silver. Economic life has become an indispensable part of everyone. Just like the Jiajing period of the Ming Dynasty, "not to divide the system but to use financial system", which fully reflects the characteristics of the times brought about by this monetary economy. Needless to say, the ability of silver to penetrate all walks of life can be glimpsed from the cultural life of the people. The Ming Dynasty literary work "Jin Ping Mei" was written during the period when silver was monetized. The economic conditions depicted in the book are unprecedentedly rich and prosperous, and for the plot of the relationship between Ximen Qing and other characters and Bai Yin can also illustrate people's reflection and criticism of the status quo at that time.

  The problem of silver hoarding was also an important factor leading to the low grain prices and low silver prices in the Ming Dynasty. However, this article classifies it as a social impact rather than a change in silver demand for analysis, because it is a social atmosphere brought about by the generalization of silver in the Ming Dynasty. There are many records of upper-level officials hoarding silver in ancient books and documents of the Ming Dynasty, and the records in the middle and late Ming Dynasty are obviously more than those in the early period. The most famous case is the house raid of the great eunuch Liu Jin during the Zhengde period. The gold and silver property seized from his home totaled 310 million taels of silver. At that time, the annual fiscal revenue of the Ming government was only 4 million taels of silver. Liu Jin's extortion in just five or six years was equivalent to the fiscal revenue of the Ming Dynasty for 75 years. The amount is amazing! Looking through the Ming Dynasty classics, there are countless similar records, which shows how bad the trend of bribery and hoarding silver was among the officials at that time. After Li Zicheng entered Beijing, he asked people to count the property in the capital, among which there were 150 million taels of silver, accounting for one-third of the silver reserves in the late Ming Dynasty. This was the case in the capital, and even more so in other regions. From high officials to wealthy merchants, everyone rushed to hide their silver. With such a huge amount of money leaving the circulation field, it is not difficult to understand why even with a huge amount of silver flowing in, the Ming Dynasty could not be dragged out of the quagmire of "money shortage". As discussed in the previous article, traditional economics believes that precious metal currency can adjust the currency stock and stabilize prices by hiding it. Marshall, the master of classical economics, also discussed the investment and emergency functions of metal currency. He believed that it was precisely because of these characteristics that metal currency could be stored for emergency use. However, these views of Western economics cannot fully explain the hoarding style of the Ming Dynasty. First, the hoarding of the Ming Dynasty was "one hoard and forever". The silver currency hidden by the nobles was not sealed for a short time, but never returned to the circulation field. This behavior was undoubtedly distorted and irrational; second, the hoarding of silver in the Ming Dynasty was more of a government behavior. Even if it was seized, the silver hoard was only transferred from the hands of high-ranking officials and wealthy merchants to the Ming Dynasty treasury. The government has long held this part of the currency, so it still cannot return to the circulation field. Under such a vicious cycle, the silver hoard in the Ming Dynasty was not a temporary self-adjustment of metal currency, but led to a permanent decline in the currency stock. Historians generally estimate that the total amount of silver flowing into the Ming Dynasty was between 200 and 300 million taels, and the silver sealed in the capital alone at the end of the Ming Dynasty was as high as 150 million taels. If we look at the whole country, the amount of silver hoarded is likely to be close to or even exceed the amount of silver flowing into the Ming Dynasty, and eventually completely offset the effect of the silver flowing in.

In summary, it is not difficult to understand why the huge influx of silver did not trigger the "price revolution" in the Ming Dynasty, nor did it promote the rapid development of the national economy of the Ming Dynasty. The Ming Dynasty was the biggest beneficiary of the great geographical discoveries in the 15th to 17th centuries, but it ultimately missed the opportunity of history. The huge influx of silver did not stimulate China in the 16th century to follow the same path as Europe, but instead gradually drifted away. Perhaps this also constitutes an explanation for the "Needham Puzzle". At the same time, it should be pointed out that there are still many deficiencies in the econometric analysis of this article. Due to the limitations of historical data and my own discipline, there are still many important indicators that are not included in the econometric model. For example, copper coins have always occupied an important position in the monetary system of the Ming Dynasty, but due to the complexity and diversity of the sources of copper coins, the stock of copper coins is difficult to count, so it is not taken into account in the model of this article; the monetization of silver triggered social changes with "fundamental social transformation nature" in the late Ming Dynasty (Wanli to Chongzhen, i.e. 1573-1644), and China's modern economic society began to take shape; the increasingly prosperous monetary economy of the Ming Dynasty gave birth to the embryonic capitalism of the late Ming society. The rapid economic development has improved people's living standards and promoted the development of culture, art, science and technology.

# 7. Conclusion and Outlook

  The monetization of silver in the Ming Dynasty not only brought prosperity to the Ming Dynasty's commerce, but also placed the Ming Dynasty's currency in the international monetary environment, and benefited the Qing Dynasty and the Republic of China, playing a positive role in the internationalization of my country's currency. The material prerequisite for silver to become the standard currency is that the silver stock is sufficient to circulate the currency. At this time, productivity needs to develop to a certain level. In the second half of the Ming Dynasty, commercial production had a certain development, the domestic market grew and expanded rapidly, commodity circulation and exchange were active, merchants in the north trafficked goods south, and merchants in the south trafficked goods north, exchanging rare items with each other. The development of commercial trade requires a stable currency to ensure the smooth and stable circulation and payment. In the Ming Dynasty, silver mines and silver bureaus were widely set up in local areas, but there was no slightly larger silver vein in my country geographically, and the output of silver was very small. For example, in July 1546, mining cost more than 30,000 gold, but only 28,500 taels of silver were obtained. Not only did it not make any profit, but it also lost the principal and wasted manpower. Such silver production could not support the huge economy of the Ming Dynasty. Due to its high price, meltability and storage, a large amount of silver circulated in the country withdrew from circulation and entered the silver cellars of corrupt officials and wealthy businessmen. Internationally, Europe fell into the dilemma of insufficient money supply due to the silver shortage in the 15th century, so Europeans were eager to find gold and silver mines around the world. The Ming Dynasty did not have the right to mint coins, and it produced very little silver. This fatal risk eventually led to rampant corruption in the Ming Dynasty and the people's livelihood was in dire straits. Internationally, silver production decreased, supply decreased, deflation occurred, the economic order of the Ming Dynasty collapsed, social life was in chaos, and it collapsed and perished. The Qing Dynasty was destroyed by a series of problems such as inflation caused by the international silver crisis. The Republic of China also caused chaos and disorder in social life because of the disorder of the currency system.

  The operation of currency has its own laws. The design of any monetary system must follow the law of currency. Relying on state power to arbitrarily promote the monetary system and ignoring its laws will inevitably lead to the failure of the monetary system. We cannot rely on state power to implement a flawed monetary system. Although money is a medium of exchange and a type of commodity, any currency used as a currency and having a payment function must take into account its own laws. Liang Fangzhong, a famous scholar, believes that although the widespread use of silver indicates the rise of the monetary economy, the negative problems that emerged after the monetization of silver are rooted in the defects of the monetary and fiscal system of the Ming Dynasty. As a weighing currency, silver in the Ming Dynasty has three types: broken silver, silver ingots and silver dollars. They are cast by silversmiths. Even if the court has supervision, the casting of silver is not uniformly implemented by the state. There are many differences in the calculation of weight and fineness. Foreign merchants therefore get business opportunities and often exchange foreign silver dollars with 90% or 70% of silver ingots of the Ming Dynasty to make profits. The monetary and fiscal system of the Ming Dynasty also violated the efficiency principle of taxation itself in fiscal science. After Zhang Juzheng's reform, there are relatively sound collection procedures such as weighing, collecting cabinets, distinguishing colors, pouring and sheathing. But in the implementation, problems will arise in every link. Officials at all levels frequently engaged in various forms of corruption, deduction and private appropriation of silver. In weighing and exchange, there were behaviors that harmed the interests of the people. Officials used the scale to cheat and seek benefits in weighing. Even if the people understood, they did not dare to approach to see the clear words, and could only swallow their anger.

  The process of silver popularization in the middle of the Ming Dynasty not only subverted the monetary system designed in the early Ming Dynasty, but also penetrated into all areas of folk production and life, triggering great changes at the social level. This great change was a chain reaction formed under the impetus of silver currency, and it was a reform of the comprehensive effects of multiple factors such as economy, politics, thought and culture. The popularization of silver brought differentiation and integration to the Ming Dynasty society: the merchant class emerged and grew, the social division of labor became more detailed, the taxation changed from in-kind and labor service to monetary tax, and the personal dependence relationship changed to economic dependence relationship. The market developed from the rural market to the urban market, the regional market and even the national market. In the late Ming Dynasty, the popularization of silver forced the Ming government to give up the right to issue the main currency, and the power of the Ming Dynasty was weakened. The demand for silver could not be balanced with the domestic silver reserves and silver mining at the time, which forced the Ming government to seek silver overseas. Overseas trade developed rapidly, and silver from Japan and America continued to flow into China. The Chinese market was able to connect with the world market, and Chinese goods circulated globally. The role of the state gave way more to the market, and the decline of dynastic power was inevitable.